\documentstyle[12pt]{article}  
\input psfig.sty

\begin{document}   
\baselineskip=17pt
\pagestyle{plain}
\setcounter{page}{1}

\begin{titlepage}
\rightline{CLNS-99/1633, NSF-ITP-99-118}
\vspace{20mm}

\begin{center}
{\huge New S-Dualities in N=2 Supersymmetric}\\
\vspace{7mm}
{\huge SU(2)\,x\,SU(2) Gauge Theory}   
\vspace{5mm}
\end{center}
\vspace{10mm}

\begin{center}
{\large Philip C. Argyres$^{1,2}$ and Alex Buchel$^{2,3}$}\\
\vspace{3mm}
{\it $^1$Newman Laboratory, Cornell University, Ithaca NY 14853\\ 
$^2$Institute for Theoretical Physics, University of California, Santa
Barbara CA 93106\\
$^3$Department of Physics, University of California, Santa Barbara CA 
93106}\\
\vspace{3mm}
{\tt argyres@mail.lns.cornell.edu, buchel@itp.ucsb.edu}
\end{center}
\vspace{15mm}

\begin{center}
{\large Abstract}
\end{center} 

\noindent   
New S-dualities in a scale invariant $N=2$ gauge theory with $SU(2)
\times SU(2)$ gauge group are derived from embeddings of the theory in
two different larger asymptotically free theories.  The true coupling
space of the scale invariant theory is a 20-fold identification of the
coupling space found in the M- and string-theory derivations of the
low energy effective action, implying a larger S-duality group.  Also,
this coupling space is different from the naively expected direct
product of two $SL(2,\bf Z)$ fundamental domains, as it contains a
different topology of fixed points.

\end{titlepage}

\newpage
\renewcommand{\baselinestretch}{1.1}  

\def\be{\begin{equation}}
\def\ee{\end{equation}}
\def\bea{\begin{eqnarray}}
\def\eea{\end{eqnarray}}

\def\bZ{{\bf Z}}
\def\bR{{\bf R}}
\def\bC{{\bf C}}
\def\CA{{\bC^2_A}}
\def\CB{{\bC^2_B}}
\def\Cf{{\bC^2_f}}
\def\bP{{\bf P}}
\def\P{{\cal P}}
\def\Q{{\cal Q}}
\def\R{{\cal R}}
\def\S{{\cal S}}
\def\T{{\cal T}}
\def\U{{\cal U}}
\def\M{{\cal M}}
\def\msu{\M_{SU}}
\def\msp{\M_{Sp}}
\def\msusu{\M_{SU\cdot SU}}
\def\msusp{\M_{SU\cdot Sp}}
\def\osp{O6${}^-$ plane}

\def\ATMP{Adv.\ Theor.\ Math.\ Phys.\ }
\def\CMP{Commun.\ Math.\ Phys.\ }
\def\JHEP{J.H.E.P.\ }
\def\NP{Nucl.\ Phys.\ }
\def\PL{Phys.\ Lett.\ }
\def\PR{Phys.\ Rev.\ }
\def\PRL{Phys.\ Rev.\ Lett.\ }


\section{Introduction and Summary}

One of the most striking elements in recent developments in our
understanding of gauge theories and string theories is the ubiquitous
appearance of S-dualities in theories with 8 or more supercharges.
S-duality denotes the exact equivalence of a theory at weak coupling
to another theory at strong coupling.  It can be described in general
as a set of identifications on the space of couplings of a theory (or
theories).  Well-known examples in four-dimensional field theory are
the $N=4$ supersymmetric Yang-Mills theories \cite{mo77} and finite
$N=2$ theories \cite{sw9408}--\cite{ll9708}.  In $N=4$ theories, for
example, S-duality identifies theories with gauge couplings $\tau$ and
$-1/\tau$.
    
The S-dualities of classes of scale invariant $N=2$ gauge theories
with simple gauge groups \cite{a9706} and product gauge groups
\cite{ab9804} were derived by embedding those theories in higher rank
asymptotically free gauge theories. The coupling space of the scale
invariant theory was realized as a submanifold of the Coulomb branch
of the asymptotically free theory.  These embedding arguments by
themselves do not necessarily capture all possible S-dualities---there
may be further identifications of the coupling space---since they only
show that a submanifold of the Coulomb branch of the appropriate
asymptotically free theory is some multiple cover of the true coupling
space.  One place where we know such further identifications must
exist are in theories with $SU(2)$ gauge group factors: for in the
limit that the other factors decouple, the remaining $SU(2)$ factor
must have the full $SL(2,\bZ)$ duality of \cite{sw9408}, rather than
the subgroup $\Gamma^0(2)\subset SL(2,\bZ)$ which emerges from the
embedding argument.  The purpose of this letter is to explore these
further S-dualities in a scale invariant $N=2$ gauge theory with
$SU(2)\times SU(2)$ gauge group.

The specific theory we focus on has massless hypermultiplets in the
representations $({\bf 2,2}) \oplus ({\bf 2,1}) \oplus ({\bf 2,1})
\oplus ({\bf 1,2}) \oplus ({\bf 1,2})$ of $SU(2)\times SU(2)$.  It has
two exactly marginal complex gauge couplings, $\tau_1$ and $\tau_2$,
which are conveniently parameterized by $f_k = e^{i\pi\tau_k}$ (so
that weak coupling is at $f_k=0$).  The new S-dualities we find act as
a 20-fold identification on $\bC^2 \simeq \{f_1,f_2\}$, and are
described explicitly in eqns.~(\ref{gampam}--\ref{fpunct}) below.  The
resulting coupling space has a single $\bZ_3$ orbifold fixed point,
complex lines of $\bZ_2$ orbifold fixed points intersecting in an
$S_3$ orbifold point, and no further strong coupling singularities.
The weak coupling singularities have the expected structure: in the
limit that one coupling vanishes, the S-duality group acts as
$SL(2,\bZ)$ on the other coupling; nevertheless, the total coupling
space is not simply the Cartesian product of two $SL(2,\bZ)$
fundamental domains.

This paper is organized as follows.  In the next section we review the
proof of the S-duality of the $SU(2)$ gauge theory \cite{a9706},
clarifying in what sense the $SL(2,\bZ)$ group of identifications on
the coupling space can be recovered.  In section 3 we study the low
energy effective action on the Coulomb branch of our scale invariant
$SU(2)\times SU(2)$ theory.  We derive two different forms of the
curve encoding this effective action by embedding the theory in either
an $SU(n) \times SU(n)$ or an $SU(2n) \times Sp(2n)$ theory.
Demanding that the resulting curves describe equivalent low energy
physics implies a non-trivial mapping between the coupling parameters
that appear in each description.  In section 4 we use this mapping and
the results of \cite{ab9804} to prove that there are the ``extra''
S-duality identifications described above.

 
\section{Deriving the SL(2,Z) duality of the SU(2) theory}

The $N=2$ theory with $SU(2) \simeq Sp(2)$ gauge group and four
massless fundamental hypermultiplets is a scale invariant theory with
an exactly marginal coupling, the complex gauge coupling $\tau$,
taking values in the classical coupling space $\M_{cl} = \{ \tau |
\hbox{Im}\tau > 0 \}$.  In \cite{sw9408} evidence was presented, in
the form of the invariance of the low energy effective action, that
the true coupling space of this theory should be the classical space
further identified under the transformations $T: \tau
\rightarrow \tau +1$ and $S: \tau \rightarrow -1/\tau$.  This gives
the coupling space as $\M = \M_{cl} / SL(2,\bZ)$,
and $SL(2,\bZ)$ is said to be the S-duality group of the
theory.\footnote{We only discuss the S-duality action on marginal
couplings and not on masses or other operators, and so will ignore the
distinction between $SL(2,\bZ)$ and $PSL(2,\bZ)$ in what follows.}

On the other hand, the duality identifications manifest in the low
energy effective action of this $SU(2)$ gauge theory derived from
either the M-theory construction of \cite{w9703} or the geometrical
engineering of \cite{kmv9706} do not comprise the full $SL(2,\bZ)$
S-duality group conjectured in \cite{sw9408}.  It was shown in
\cite{a9706} that the true coupling space of the scale invariant
$SU(2)$ gauge theory can be derived from its different covering spaces
represented by submanifolds of Coulomb branches of two different
embeddings of this theory in higher rank asymptotically free
theories. In this section we review this argument and clarify the
relation between the geometry of the covering of the coupling space
and the S-duality group.

Consider first the scale invariant $SU(2)$ theory with four massless
hypermultiplets in the fundamental representation. The Coulomb branch
of the theory is described by the curve \cite{sw9408,aps9505}
\be
y^2=(v^2-u)^2-4 f v^4,
\label{su2}
\ee
parameterized by the gauge coupling $f$ and the gauge invariant
adjoint vev $u$, a local coordinate on the Coulomb branch.  $f$ is a
function of the coupling such that $f\sim e^{i\pi\tau}$ at weak
coupling.\footnote{In the $N=2$ theories discussed here it is
convenient to define the coupling as $\tau = {\vartheta \over \pi} + i
{8 \pi \over g^2}$, differing by a factor of two from the usual
definition.}  Embedding this theory into the asymptotically free
$SU(3)$ model with $4$ quarks and scaling on the Coulomb branch of the
latter (while tuning appropriately the masses of the quarks) to the
scale invariant $SU(2)$ theory one identifies \cite{a9706} the
coupling space $\msu=\{f\}$ with $\bP^1$ with two punctures and an
orbifold point: a weak coupling singularity $f=0$, an ``ultra-strong''
coupling point at $f=1/4$, and a $\bZ_2$ orbifold singularity at
$f=\infty$.

On the other hand, this scale invariant $N=2$ $SU(2)$ gauge theory can
be thought of as an $Sp(2)$ theory with $4$ massless fundamental
flavors, whose curve reads \cite{as9509}
\be
y^2=x (x-v)^2-4g x^3.
\label{sp2}
\ee
The coupling space $\msp=\{g\}$ of this theory was derived in
\cite{a9706} from its embedding in asymptotically free $Sp(4)$ theory
with $4$ massless hypermultiplets by tuning on the Coulomb branch of
the latter to the scale invariant $Sp(2)$ theory. One then finds that
$\msp$ is again the complex manifold $\bP^1$ with two punctures and an
orbifold point: a weak coupling singularity at $g=0$, an
``ultra-strong'' singularity at $g=1/4$, and a $\bZ_2$ orbifold
singularity at $g=\infty$.

Both the $SU(2)$ and the $Sp(2)$ descriptions of the scale invariant
theory must describe the same physics. In particular, their low energy
effective actions described by the complex structure of the curves
(\ref{su2}) and (\ref{sp2}) must be the same.  We therefore look for
an $SL(2,\bC)$ transformation on $x$ which maps the zeros of the right
sides to one another.  Of the six distinct such mappings only two map
weak coupling to weak coupling, and imply the identification
\be
f={4\sqrt{g}(1+2\sqrt{g})\over (1+6\sqrt{g})^2}.
\label{fg}
\ee
Choosing different signs of the square root gives two maps between
$\msp$ and $\msu$, which induce the nontrivial identification $\U$ on
$\msu$ 
\be\label{iden}
\U(f) = {\gamma + 2 \over (\gamma+3)^2}
\ee
where $\gamma$ is a root of
\be\label{iden2}
0 = f \gamma^2 + \gamma +
1 .
\ee 
This gives two maps from $\msu$ to itself, one for each $\gamma$
satisfying (\ref{iden2}).  Thus these identifications imply at least a
three-fold identification on $\msu$ (the original point and its two
images).  In fact, a little algebra shows that the orbit of a generic
point under $\U$ is just this set of three points, so $\msu$ is a
triple cover of the true coupling space of the scale invariant $SU(2)$
theory.  In particular, the identifications (\ref{iden}) map the
``strong coupling'' point $f=1/4$ to the $f=0$ weak coupling
singularity, and map the $\bZ_2$ point $f=\infty$ to the point
$f=2/9$.  In addition, there is a new fixed point under these
identifications, namely $f=1/3$, which it is easy to check is a
$\bZ_3$ orbifold point.  The net result is that with these further
identifications, the coupling space becomes topologically a sphere
with three special points: the weak coupling puncture (the image of
$f=0$ or $1/4$), a $\bZ_2$ orbifold point (the image of $f=2/9$ or
$\infty$), and a $\bZ_3$ orbifold point (the image of $f=1/3$).  Since
the map (\ref{iden}) is analytic, the true coupling space inherits a
complex structure from that of the punctured $f$-sphere.  The order of
the orbifold points reflects the nature of the singularity in the
complex structure at the punctures.

This argument shows that there are indeed more identifications on the
coupling space than were apparent in either the $SU(2)$ form of the
curve (\ref{su2}) or the $Sp(2)$ form of the curve (\ref{sp2}).  But
it might not be clear from this argument how to actually see the
$SL(2,\bZ)$ structure of the duality group.  For this we need an
intrinsic definition of what we mean by duality group.  Since having
an S-duality group $\Gamma$ means that the coupling space is given by
$\M = \M_{cl}/\Gamma$, and the classical coupling space $\M_{cl}$ is
simply connected, we can define \cite{w9703}
\be
\Gamma = \pi_1(\M).
\ee
When $\M$ has orbifold singularities, $\pi_1(\M)$ should be
understood in the orbifold sense \cite{w9703}, meaning that the
generator $U$ of $\pi_1(\M)$ corresponding to looping about a $\bZ_n$
orbifold point satisfies $U^n=1$.  

The true $SU(2)$ coupling space deduced above has the complex
structure of a sphere with one puncture, a $\bZ_2$ orbifold point,
and a $\bZ_3$ orbifold point.  Thus the S-duality group $\pi_1(\M)$
has two generators which we can take to be $U$, generating loops
around the $\bZ_2$ point, and $V$, generating loops around the $\bZ_3$
point, and satisfying $U^2=V^3=1$.  There are no other constraints
since we know that going around the weak coupling puncture is a
$\theta$-angle rotation, which does not correspond to any orbifold
identification.  But $SL(2,\bZ)$, considered as an abstract infinite
discrete group, can be presented as the group with two generators $S$
and $T$ satisfying only the relations $S^2 = (ST)^3 = 1$.  So,
identifying $S=U$ and $ST=V$, we see that the S-duality group is
isomorphic to $SL(2,\bZ)$.


\section{Curves for the SU(2)\,x\,SU(2) theory}

In preparation for our discussion of S-duality in the $SU(2)\times
SU(2)$ scale invariant theory, we must first make a somewhat lengthy
technical detour to derive useful forms for the curves whose complex
structure encodes the low energy physics of the Coulomb branch of the
theory.  The different curves we need are those arising from viewing
the $SU(2)\times SU(2)$ theory as part of an $SU(n)\times SU(n)$
series or as part of an $SU(2n)\times Sp(2n)$ series.  The goal of
this section is to derive an explicit map between the couplings of the
two versions of the theory---the analog of eq.~(\ref{fg}) above.  This
map is summarized at the end of this section for those who prefer to
skip the technicalities.

We start by briefly reviewing the derivation \cite{w9703} of the
$SU\times SU$ curves from an M5 brane configuration in M-theory.  In
subsection 3.2 we then derive curves for the $SU\times Sp$ series with
fundamental matter using an M5 brane configuration on $\bR^7 \times Q$
where $Q$ is the Atiyah-Hitchin manifold, corresponding to a
negatively charged O6 orientifold in a type IIA string picture.  In
subsection 3.3 we specialize to vanishing bare masses for the matter
hypermultiplets in the $SU(2)\times SU(2)$ and $SU(2)\times Sp(2)$
curves, develop hyperelliptic forms for both curves, and then derive
the mapping of parameters matching the two.  In subsection 3.4 we
summarize the results of this section relevant for our discussion of
S-duality.

\subsection{SU\,x\,SU curves}

Consider the scale invariant $SU(n)\times SU(n)$ theory with one
hypermultiplet in the bifundamental, $n$ in the first $SU(n)$
fundamental, and $n$ in the second $SU(n)$ fundamental.  This can be
realized as a IIA brane configuration by placing three NS5 branes
along the $x^{0\cdots5}$ directions separated in $x^6$ but located at
equal values in $x^{7\cdots9}$, and $n$ D4 branes along the
$x^{0\cdots3}$ and $x^6$ directions suspended between neighboring
pairs of NS5 branes.  The fundamental matter is incorporated by
including $n$ semi-infinite D4 branes extending to the right and left
in the $x^6$ direction.

It is easy to lift such a brane configuration to an M-theory
curve \cite{w9703}
\be
F(t,v) \equiv p(v) t^3 + q(v) t^2 + r(v) t + s(v) = 0,
\label{susucrv}
\ee
where $v=x^4+ix^5$, $t=\exp\{(x^6+ix^{10})/R\}$, $x^{10}$ is the
eleventh dimension of radius $R$.  $p$, $q$, $r$ and $s$ are
polynomials of degree $n$:
\bea
p &=& \prod_{i=1}^n (v-m^{(1)}_i-M), \cr
q &=& B_1 \cdot \prod_{j=1}^n (v-b_j-M), \cr
r &=& B_2 \cdot \prod_{k=1}^n (v-a_k+M), \cr
s &=& \prod_{\ell=1}^n (v-m^{(2)}_\ell+M).
\label{psucoeff}
\eea
The leading coefficients of $p$ and $s$ are set to $1$ by rescaling
$t$ and $v$, and by a shift in $v$ we set $\sum_k a_k = \sum_j b_j =0$.
Interpreting the positions in the $v$ plane of the D4 branes as
mass parameters or Coulomb branch vevs, we find that the $m^{(1)}_i$
and $m^{(2)}_\ell$ are the bare masses of the  fundamentals of the first and
the second $SU$ factors, $M$ is the bifundamental mass, and the traceless
$a_k$ and $b_j$ are the eigenvalues of the adjoint vevs of the first 
and the second $SU$ factors.

The $B_i$ in (\ref{psucoeff}) encode the gauge couplings through the
relative asymptotic positions of the NS5 branes in the IIA picture.
These positions are given by the roots of $F(t,v)=0$ for large $v$,
that is, the roots of $t^3 + B_1 t^2 + B_2 t + 1 = 0$.  The relative
positions of these roots are unaffected by the $\bZ_3$ transformation
of the coefficients $B_i$
\be
(B_1,B_2)\to(\omega^p B_1,\omega^{2p} B_2)\qquad p=1,2,
\label{z3}
\ee
where $\omega=e^{2\pi i/3}$.  Thus the space $\msusu$ of inequivalent
couplings that enters into the low-energy physics on the Coulomb
branch of this $SU(n)\times SU(n)$ theory is the space $\bC^2 \simeq
\{B_1, B_2\}$ modded by the $\bZ_3$ action (\ref{z3}).  Furthermore,
in addition to the $\bZ_3$ orbifold fixed point at $B_1=B_2=0$, this
space has singularities whenever the asymptotic positions of the
M5 branes collide---whenever $0 = 27 - 18 B_1 B_2 - B_1^2 B_2^2 + 4
B_1^3 + 4 B_2^3$---as well as weak-coupling singularities whenever one
of the NS5 branes goes off to infinity: $B_1 \to \infty$ or $B_2 \to
\infty$.  Indeed, the space of $SU\times SU$ couplings can be
parameterized by the $\bZ_3$-invariant combinations $f_1 \equiv
B_1/B_2^2$ and $f_2
\equiv B_2/B_1^2$, which have been chosen to correspond to the
normalization of the $SU(2)$ coupling $f$ used in (\ref{su2}), so that
they are related to gauge couplings at weak coupling as
$\{f_1,f_2\}\sim \{e^{i\pi\tau_1}, e^{i\pi\tau_2} \}$.  

We can check this identification of the coupling parameters (as well
as our implicit identification of the vevs and bare masses in the
$SU\times SU$ curve) by decoupling one of the $SU$ factors by taking
one of the NS5 branes off to infinity.  For example, we can decouple
the first $SU$ factor by setting $B_2 = f_2 B_1^2$ with $f_2$ finite,
and sending $B_1 \to \infty$.  The $SU\times SU$ curve (\ref{susucrv})
then becomes, after rescaling $t \to B_1 t$ and dividing by $B_1^3$,
\be
0 = t\left( p(v) t^2 + {q(v)\over B_1} t + {r(v)\over B_1^2} \right).
\ee
The overall factor of $t$ is for the decoupled brane, and the
remaining polynomial becomes, using (\ref{psucoeff}),
\be
0 = \prod_{i=1}^n (v-m^{(1)}_i-M) \cdot t^2 + 
\prod_{j=1}^n (v-b_j-M) \cdot t + f_2 \cdot \prod_{k=1}^n (v-a_k+M).
\ee
Multiplying by $\prod_{i=1}^n(v-m^{(1)}_i-M)$, changing variables to
$y= 2t\prod_{i=1}^n (v-m^{(1)}_i-M) + \prod_{j=1}^n (v-b_j-M)$,
shifting $v\to v+M$, and identifying $M_i = m^{(1)}_i$ for
$i=1,\ldots,n$ and $M_i = a_{i-n}-2M$ for $i=n+1,\ldots,2n$, gives the
scale invariant $SU$ curve found in \cite{aps9505}.

\subsection{SU\,x\,Sp curves}

Consider the scale invariant $SU(2n)\times Sp(2n)$ theory with one
hypermultiplet in the bifundamental, $2n$ in the $SU(2n)$ fundamental,
and 2 in the $Sp(2n)$ fundamental.  This can be realized as a IIA
brane configuration in the presence of an O6 orientifold plane of
negative RR charge \cite{ll9708}.  The \osp\ is the fixed point of a
$\bZ_2$ quotient which acts on the space-time coordinates as
$x^{4,5,6} \rightarrow -x^{4,5,6}$, and thus extends along the
$x^{0\cdots3}$ and $x^{7\cdots9}$ directions, and is located at
$x^{4\cdots6}=0$.  It is convenient to work on the double cover, by
including mirror images for all branes, where the \osp\ has RR charge
-8 in D6 brane units.  The $SU(2n) \times Sp(2n)$ gauge theory is then
constructed by placing two NS5 branes (and their mirror images) along
the $x^{0\cdots5}$ directions separated in $x^6$ but located at equal
values in $x^{7\cdots9}$, and $2n$ D4 branes along the $x^{0\cdots3}$
and $x^6$ directions suspended between neighboring pairs of
NS5 branes.  The fundamental matter is incorporated by including
D6 branes parallel to the \osp: two between the \osp\ and the first
NS5 brane, and $2n$ between the two NS5 branes (as well as their
mirror images).

Following \cite{u9803}, we can derive the curve for this brane
configuration by first moving the D6 branes to left and right
infinity, whereupon they drag D4 branes behind them upon passing
through any NS5 branes \cite{hw9611}.  Also, we can represent the
\osp\ as a ``neutral'' O6 plane by pulling in 2 D6 branes (and their
mirror images) from infinity to cancel the \osp\ RR charge.  Upon
passing through the NS5 branes, the D6 branes create two D4 branes
between the NS5 branes and four between the NS5 brane and the \osp\
(as well as their mirror images).  Thus the final configuration is
simply four NS5 branes crossed by $2n+4$ infinite D4 branes, all
arranged symmetrically with respect to the origin of $x^{4\cdots6}$.

It is easy to lift such a brane configuration to the M-theory curve
\be
F(t,v) \equiv p(v) t^4 + q(v) t^3 + r(v) t^2 + q(-v) t + p(-v) = 0
\label{suspcrv}
\ee
where $v=x^4+ix^5$, $t=\exp\{(x^6+ix^{10})/R\}$, $x^{10}$ is the
eleventh dimension of radius $R$; $p$, $q$ and $r$ are polynomials of
degree $2n+4$, $r(v)=r(-v)$, and the $\bZ_2$ identification is lifted
to
\be\label{orbif}
(v,t) \to (-v,1/t).
\ee
The condition that there be an \osp\ implies \cite{ll9708} that this
curve be non-singular on the Atiyah-Hitchin space.  As discussed in
\cite{u9803}, this in turns implies that $(\partial^\ell F/\partial 
v^\ell)|_{v=0}$ has a zero of order $4-\ell$ at $t=-1$ for
$\ell=0,\ldots,3$, giving
\bea
p &=& \prod_{i=1}^2 (v-m_i)^2 \cdot
\prod_{j=1}^{2n} (v-\mu_j-M), \cr
q &=& 4 p[0] + 2 v p[1] + A_1 \cdot v^2 \cdot
\prod_{i=1}^2 (v-m_i) \cdot
\prod_{k=1}^{2n} (v-a_k-M), \cr
r &=& 6 p[0] + 2 v^2 (q[2]-p[2]) + A_2 \cdot v^4 
\cdot \prod_{\ell=1}^n (v^2 - b_\ell^2) ,
\label{pspcoeff}
\eea
where $p[n]$ refers to the coefficient of $v^n$ in $p(v)$.
Interpreting the positions in the $v$ plane of the D4 branes as mass
parameters or Coulomb branch vevs, we find that the $m_i$ are the bare
masses of the two $Sp$ fundamentals, $\mu_j$ are the masses of the
$SU$ fundamentals, $M$ is the bifundamental mass, the traceless $a_k$
are the eigenvalues of the $SU$ adjoint vev, and the $b_\ell$ likewise
for the $Sp$ adjoint vev.

The $A_i$ in (\ref{pspcoeff}) encode the gauge couplings through the
relative asymptotic positions of the NS5 branes in the IIA picture.
These positions are given by the roots of $F(t,v)=0$ for large $v$,
that is, the roots of $t^4 + A_1 t^3 + A_2 t^2 + A_1 t + 1 = 0$.  The
relative positions of these roots are unaffected by the $\bZ_2$
transformation of the $A_i$ coefficients
\be
(A_1, A_2) \to (-A_1, A_2).
\label{z2}
\ee 
Thus the space $\msusp$ of inequivalent couplings that enters into
the low-energy physics on the Coulomb branch of this $SU(2n)\times
Sp(2n)$ theory is the space $\bC^2 \simeq \{A_1, A_2\}$ modded by the
$\bZ_2$ action (\ref{z2}).  Furthermore, in addition to the line of
$\bZ_2$ orbifold fixed points at $A_1=0$, this space has strong
coupling singularities whenever the asymptotic positions of the M5
branes collide, which is when $A_2 + 2 = \pm 2 A_1$ or $A_1^2 = 4A_2 -
8$, as well as weak coupling singularities whenever one of the M5
branes goes off to infinity: $A_1 \to \infty$ or $A_2 \to \infty$.
Indeed, the space of $SU\times Sp$ couplings can be parameterized by
the $\bZ_2$-invariant combinations $g_1 \equiv A_2/A_1^2$ and $g_2
\equiv A_1^2/A_2^2$, which have been chosen to correspond to the
normalization of the $SU$ and $Sp$ couplings used in the last section,
so that they are related to gauge couplings at weak coupling as
$\{g_1,g_2\}\sim \{e^{i\pi\tau_1}, e^{i\pi\tau_2} \}$ where $\tau_1$
is the $SU$ coupling and $\tau_2$ the $Sp$ coupling.

We can check this identification of the coupling parameters (as well
as our implicit identification of the vevs and bare masses) in the
$SU\times Sp$ curve by decoupling the $Sp$ factor ($g_1$ fixed, $A_i
\to \infty$) or the $SU$ factor ($g_2$ fixed, $A_i \to \infty$).
Decoupling the $Sp$ factor leads to considerations very similar to
those discussed above in the case of the $SU\times SU$ curve, so we
consider only the decoupling of the $SU$ factor.  This decoupling is
also interesting since it involves passing from the $\{v,t\}$ space
which is a double cover of the orbifold space, to the single-valued
coordinates which resolve the orbifold singularity appropriately.  We
will need to do the same change of variables on the $SU(2)\times
Sp(2)$ curve in the next subsection.

The $SU\times Sp$ curve (\ref{suspcrv}) then becomes
\be
0 = t\left( q(v) t^2 + r(v) t + q(-v)\right).
\ee
The overall factor of $t$ is for the decoupled brane, and the
remaining polynomial becomes
\be
0 = \sqrt{g_2} \prod_{i=1}^{2n+2}(v-M_i) \cdot t
+ \left[2\sqrt{g_2} \prod_{i=1}^{2n+2} M_i
+ v^2 \prod_{\ell=1}^n(v^2-b_\ell^2)\right] 
+ \sqrt{g_2} \prod_{i=1}^{2n+2}(v+M_i) \cdot {1\over t},
\label{spsccrv}
\ee
where we have used (\ref{pspcoeff}), divided by $A_2 v^2/t$, and
defined $M_i = m_i$ for $i=1,2$ and $M_i = a_{i-2}$ for
$i=3,\ldots,2n+2$.  In order to compare this curve with previously
derived genus-$n$ $Sp(2n)$ curves, we must divide out the orbifold
identifications (\ref{orbif}).  To do this, define the invariant
coordinates
\bea\label{invcoords}
x &=& v^2\nonumber\\
y &=& [t - (1/t)]v^{-1} \nonumber\\
z &=& [t + (1/t) +2]v^{-2},
\eea
which are related by
\be\label{athi}
y^2 = x z^2 - 4 z.
\ee
Note that the change of variables (\ref{invcoords}) is singular when
$v=0$; it serves to resolve the orbifold singularities at $v=0$,
$t=\pm 1$ so that the resulting space has the complex structure of the
Atiyah-Hitchin space \cite{AH}, which is the appropriate M-theory
resolution of the \osp\ singularity \cite{SSW}.  In these variables,
the curve (\ref{spsccrv}) becomes
\be\label{spintcrv}
0 = xP_0(x) \cdot z + xP_1(x) \cdot y -2P_0(x)
+ 2\sqrt{g_2} \prod_{i=1}^{2n+2} M_i 
+ x \prod_{\ell=1}^n(x-b_\ell^2),
\ee
where we have defined $P_0$ and $P_1$ by $\sqrt{g_2}
\prod_{i=1}^{2n+2} (v-M_i) = P_0(v^2) + vP_1(v^2)$.  Making the
change of variables $\widetilde y = P_0 y + xP_1 z -2P_1$ and
$\widetilde z = xP_1 y + xP_0 z -2P_0$, (\ref{athi}) and
(\ref{spintcrv}) become
\bea\label{spint2}
\widetilde z &=& -2\sqrt{g_2} \prod_{i=1}^{2n+2} M_i - x 
\prod_{\ell=1}^n(x-b_\ell^2), \nonumber\\
x{\widetilde y}^2 &=& {\widetilde z}^2 
- 4 g_2 \prod_{i=1}^{2n+2}(x-M_i^2),
\eea
where we have used the identity $P_0^2 - x P_1^2 = g_2
\prod_{i=1}^{2n+2} (x-M_i^2)$.  Eliminating $\widetilde z$ in
(\ref{spint2}) then gives the $Sp(2n)$ curve found in 
\cite{as9509}.

\subsection{SU(2) x SU(2) and SU(2) x Sp(2) scale invariant curves}

We now specialize to the $SU(2)\times SU(2)$ and $SU(2)\times
Sp(2)$ theories which are of interest for the S-duality argument.

Consider first the $SU(2)\times SU(2)$ scale invariant theory with
zero bare masses for the hypermultiplets.  {}From (\ref{susucrv},
\ref{psucoeff}) the Coulomb branch of this theory is described by
\be\label{su2c}
t^3 v^2+B_1 t^2 (v^2-u_1)+B_2 t (v^2-u_2)+v^2=0,
\ee
where $u_1=-b_1 b_2$ and $u_2=-a_1 a_2$ denote the Coulomb branch
moduli of the two $SU(2)$'s.  To study degenerations of (\ref{su2c})
on the Coulomb branch it is convenient to represent it as a double
cover of the complex $t$ plane:
\be\label{planesu}
v^2={t (B_1 u_1 t + B_2 u_2) \over (t^3+B_1 t^2+B_2 t+1)}.
\ee 
The change of variables
\be\label{chova}
y = (t^3 + B_1 t^2 + B_2 t + 1) v
\ee
takes (\ref{planesu}) to the hyperelliptic form
\be\label{hypersu}
y^2=t (B_1 u_1 t + B_2 u_2) (t^3+B_1 t^2+B_2 t+1).
\ee

We pause here to discuss the validity of changes of variables like
(\ref{chova}), which we will use again below on the $SU(2)\times
Sp(2)$ curve, and which we also used in the decoupling checks of the
last subsections.  It is important that the complex structures of
curves related by these changes of variables are the same since we
will match the parameters of the $SU(2)\times SU(2)$ and $SU(2)\times
Sp(2)$ curves by comparing the complex structures of their
hyperelliptic forms.  The issue is the apparent singularity of the
change of variables (\ref{chova}) whenever $t^3 + B_1 t^2 + B_2 t +
1=0$.  In fact this change of variables, when properly understood, is
not singular on the curve, and so the resulting hyperelliptic curve
(\ref{hypersu}) is equivalent to (has the same complex structure as)
the prior curve (\ref{planesu}).

The key point lies in the treatment of the points at infinity on the
curves.  Let us generalize to a situation where we have a curve of the
form
\be\label{plncurve}
v^2 \prod_{j=1}^m (t-f_j)= \prod_{i=1}^m (t-e_i),
\ee
which we would like to think of as representing a Riemann surface of
genus $m-1$.  Thought of as a curve embedded in $\bC^2 = \{v,t\}$,
though, (\ref{plncurve}) is non-compact, going off to infinity as $t
\to f_j$ and $t\to \infty$.  We can compactify this curve by replacing
the $\{v,t\}$ space with an appropriate projective space; the correct
choice of projective space is determined by demanding that the genus
of the resulting compact surface indeed be $m-1$.  This is achieved if
each infinity $t\to f_i$ is replaced by a single point, while the
$t\to\infty$ infinity is compactified at two distinct points.  The
appropriate projective space which does this is the direct product of
two Riemann spheres, $\bP^1 \times \bP^1$, which can be defined as
$\bC^4 = \{u,v,s,t\}$ modulo the identifications $\{u,v,s,t\} \simeq
\{ u,v, \lambda s, \lambda t\}$ for $\lambda \in \bC^*$, and 
$\{u,v,x,z\} \simeq \{ \mu u, \mu v, s, t\}$ for $\mu\in\bC^*$.  The
curve is homogenized to $v^2 \prod_{j=1}^m (t-f_js )= u^2
\prod_{i=1}^m (t-e_i s)$.  The infinities of the $\{v,t\} = \bC^2$ 
space are compactified to two (intersecting) copies of $\bP^1$ in
$\bP^1 \times \bP^1$, while the homogeneous curve intersects these
``infinities'' at the points $\{u,v,s,t\} = \{0,1,1,f_j\}$
(corresponding to $t\to f_j$) and $\{1,\pm1,0,1\}$ (corresponding to
$t\to \infty$).

We are interested in the change of variables $y=v\cdot\prod_{j=1}^m
(t-f_j)$ which in homogeneous coordinates can be written $y = (v/u)
\cdot \prod_{j=1}^m (t-f_j s)$, $x = t$, $z = s$.  The $\bP^1
\times \bP^1$ identifications on $\{u,v,s,t\}$ imply $\{y,x,z\}
\simeq \{ \lambda^m y, \lambda x, \lambda z\}$ for $\lambda \in
\bC^*$, which defines a point in the weighted projective space ${\bf
P}^2_{(m,1,1)}$.  This is a smooth space except for a $\bZ_m$ orbifold
singularity at the point $\{y,x,z\} = \{1,0,0\}$.  The change of
variables thought of as a map from $\bP^1 \times \bP^1 \to
\bP^2_{(m,1,1)}$ is singular on the $\bP^1$ at $v=\infty$ which is
mapped to the $\bZ_m$ orbifold point of ${\bf P}^2_{(m,1,1)}$, except
for the points $\{v,t\} = \{ \infty, f_i\}$ which are blown up to the
$\bP^1$ of points $\{y,x,z\} = \{ {\cdot}, f_i, 1 \}$.

The image of the homogeneous curve under this change of variables
is the genus $m-1$ hyperelliptic curve
\be\label{hypcurve}
y^2 = \prod_{i=1}^m (x-e_iz) (x-f_iz),
\ee
which does not intersect the $\bZ_m$ orbifold point of ${\bf
P}^2_{(m,1,1)}$ if $\prod_i e_i f_i \neq 0$.  In particular, the ${\bf
P}^1 \times \bP^1$ curve approaches the points $\{u,v,s,t\} =
\{0,1,1,f_j\}$ in such a way that their images in ${\bf
P}^2_{(m,1,1)}$ miss the orbifold point.  Therefore the change of
variables is a holomorphic mapping between the abstract Riemann
surfaces, and so equates their complex structures.

In the case of the $SU(2)\times SU(2)$ curve (\ref{planesu}) the $f_i$
are roots of $t^3+B_1 t^2+B_2 t+1=0$, while the $e_i$ are $0$, $-(B_2
u_2)/(B_1 u_1)$, and $\infty$.  The branch points at zero and infinity
are harmless as can be seen by the fact that an $SL(2,\bC)$
transformation on the $\{s,t\}$ $\bP^1$ preserves the complex
structure of the curve and can be used to move all branch points to
finite points on the $t$ plane.

We return now to discuss the $SU(2)\times Sp(2)$ theory.  {}From
(\ref{suspcrv}) and (\ref{pspcoeff}) the curve of the scale invariant
$SU(2)\times Sp(2)$ theory with zero hypermultiplet masses is given by
\be
v^2 t^4+A_1 (v^2-v_1) t^3+A_2 (v^2-v_2) t^2+ A_1 (v^2-v_1) t+v^2=0,
\label{sp2c}
\ee
where $v_1=-a_1 a_2$ and $v_2=b_1^2$ are Coulomb branch moduli of the
$SU$ and $Sp$ factor respectively.  This curve is of the form
(\ref{plncurve}) with $m=4$ (and one $f_j$ at infinity), thus
describing a genus 3 Riemann surface (as is also clear from its brane
construction).  It was supposed to be equivalent to the $SU(2)\times
SU(2)$ curve, which was genus 2.  The reason for the mismatch is that
the $SU(2)\times Sp(2)$ curve was constructed on the double cover of
the \osp\ orbifold space.

Changing to single-valued variables on the orbifold space {\it via}
(\ref{invcoords}), which parameterize the non-singular Atiyah-Hitchin
space \cite{AH} (the M theory resolution of the space transverse to
the \osp\ \cite{SSW}), gives the curve (\ref{sp2c}) as the
intersection of the surfaces
\bea
y^2 &=& xz^2 - 4z,\nonumber\\
0 &=& x ((xz-2)^2-2) + A_1 (x-v_1) (xz-2) + A_2 (x-v_2).
\eea
Change variables by $s=xz-2$, leaving $x$ and $y$ unchanged.  Then the
curve becomes the intersection
\bea\label{intersect}
x y^2 &=& s^2 - 4,\nonumber\\
0 &=& x (s^2-2) + A_1 (x-v_1) s + A_2 (x-v_2).
\eea
This change of variables is singular at $x=0$ which is a
direction at infinity on the curve.  As in the discussion above,
as long as we treat the ``points'' at infinity correctly so
as to preserve the genus of the curve, the complex structure
will be preserved by the change of variables.
Eliminating $x$ from (\ref{intersect}) gives the curve
\be\label{planesp}
y^2 = {(s^2 - 4) (s^2+ A_1 s + A_2 -2) 
\over (A_1 v_1 s + A_2 v_2)} .
\ee
($x$ was the right variable to eliminate since only $x$ is single
valued on the Atiyah-Hitchin space, which is a double cover of the
$y$-$z$ plane.)  Finally, by the type of change of variables discussed
above, $w = (A_1 v_1 s + A_2 v_2) y$, the genus 2 curve emerges in the
hyperelliptic form
\be\label{hypersp}
w^2 = (A_1 v_1 s + A_2 v_2) (s^2 - 4) (s^2+ A_1 s + A_2 -2).
\ee

Since the $SU(2)\times SU(2)$ and $SU(2)\times Sp(2)$ theories are
physically identical, the two genus 2 hyperelliptic curves
(\ref{hypersu}) and (\ref{hypersp}) must have the same complex
structure as a function of the couplings and vevs.  Thus there must be
an $SL(2,\bC)$ transformation relating $t$ and $s$ which maps the
branch points of (\ref{hypersu}) to those of (\ref{hypersp}).  If we
map the branch points at infinity to each other, and the branch point
at $s=-2$ to the one at $t=0$, then we must find a linear
transformation $4\beta t= s+2$ and a map between the vevs and
couplings which satisfies
\be
(A_1 v_1 s + A_2 v_2) (s - 2) (s^2 + A_1 s + A_2 -2) \propto
(B_1 u_1 t + B_2 u_2) (t^3 + B_1 t^2 + B_2 t + 1),
\ee
for some $\beta$.  Since the theory is scale-invariant, we can choose
an arbitrary relative scaling of the $u$ and $v$ vevs so that $u_1 =
v_1$.  We then find the following relations between couplings,
\bea\label{cmatch}
A_1 &=& 8 + 4\beta B_1,\nonumber\\
A_2 &=& 30 + 24\beta B_1 + 16\beta^2 B_2,\nonumber\\
0 &=& 1 + \beta B_1 + \beta^2 B_2 + \beta^3 ,
\eea
while the vevs are related by $v_1 = u_1$ and $(A_2/A_1) v_2 = 2u_1 +
4 \beta (B_2/B_1) u_2$.  These matching relations are the main result
of this section.  They can be inverted to read
\bea\label{cmatchinv}
B_1 &=& (A_1-8)/(4\alpha),\nonumber\\
B_2 &=& (A_2-6A_1+18)/(16\alpha^2),\nonumber\\
16\alpha^3 &=& 2 A_1 - A_2 - 2,
\eea
for the couplings, with the vevs related by $u_1 = v_1$ and
$(B_2/B_1) u_2 = [(A_2/A_1) v_2 - 2 v_1 ]/(4\alpha)$, corresponding
to a map $4\alpha t = s+2$ between the curves.

Finally, one can easily check that in the weak coupling limits, the
above matching of parameters reduces to the appropriate
identifications.  For example, decoupling the $SU(2)$ factor of the
$SU(2)\times Sp(2)$ theory by sending $A_i\to\infty$ keeping $g_2=
A_1^2/A_2^2$ fixed (and thus $g_1 \to 0$), (\ref{cmatchinv}) implies
that the $SU(2)\times SU(2)$ couplings go as
\bea\label{checkslz}
f_1 \equiv {B_1\over B_2^2} &\to& {4\sqrt{g_2}(1+2\sqrt{g_2})\over 
(1+6\sqrt{g_2})^2}, \nonumber\\
f_2 \equiv {B_2\over B_1^2} &\to& 0,
\eea
which recovers precisely the mapping (\ref{iden}) between the $SU(2)$
and $Sp(2)$ couplings used in section 2, and is a non-trivial
consistency check on the calculations of this section.

\subsection{Summary of SU(2)\,x\,SU(2) low energy coupling spaces}

We now summarize what we have just derived about the space of
couplings of the $SU(2)\times SU(2)$ theory as they appear in the low
energy effective actions on the Coulomb branch described by the $SU(2)
\times SU(2)$ and $SU(2) \times Sp(2)$ curves.  We denote these two 
spaces of couplings by $\msusu$ and $\msusp$ respectively.

\subsubsection{$\msusu$}

The $SU(2)\times SU(2)$ low energy effective action is described by
two complex couplings $B_1$ and $B_2$ which parameterize an $\msusu
\simeq \bC^2/S_3$ orbifold space.  The $S_3$ orbifold identifications
are generated by the $\bZ_3$ element
\be
\P: (B_1, B_2) \to (\omega B_1, \omega^2 B_2),
\label{z3sum}
\ee
where $\omega$ is a cube root of unity, as well as by the $\bZ_2$
element
\be
\Q: (B_1, B_2) \to (B_2, B_1)
\label{sutosum}
\ee
which simply interchanges the two $SU(2)$ factors.  Resulting from the
$S_3$ identifications, $\msusu$ has three lines of $\bZ_2$ orbifold
singularities when $B_1=\omega B_2$ which intersect in an $S_3$
orbifold point at $B_1=B_2=0$.  $\msusu$ also has strong-coupling
singularities when
\be\label{susings}
0 = 27 - 18 B_1 B_2 - B_1^2 B_2^2  + 4 B_1^3 + 4 B_2^3
\ee
as well as weak-coupling singularities when $B_1 \to \infty$ or $B_2
\to \infty$.  The $\bZ_3$-invariant couplings
\be
f_1\equiv{B_1\over B_2^2}\qquad {\rm and}\qquad
f_2\equiv{B_2\over B_1^2},
\label{fsum}
\ee
are related to the $\{\tau_1,\tau_2\}$ gauge couplings of the
two $SU(2)$ factors by $\{f_1,f_2\}\sim \{e^{i\pi\tau_1},
e^{i\pi\tau_2} \}$ at weak coupling.  

\subsubsection{$\msusp$}

The $SU(2) \times Sp(2)$ curve, though describing the same theory, has
a very different space of couplings, $A_1$ and $A_2$, parameterizing
the orbifold space $\msusp \simeq \bC^2/\bZ_2$.  The $\bZ_2$
identification acts as
\be
\R: (A_1,A_2) \to (-A_1, A_2),
\label{z2sum}
\ee
and gives rise to a line of $\bZ_2$ orbifold fixed points in $\msusp$
when $A_1=0$.  In addition, $\msusp$ has strong coupling singularities
when
\be\label{spsings}
A_2 + 2 = \pm 2 A_1 \qquad\mbox{or}\qquad A_1^2 = 4A_2 - 8,
\ee
as well as weak-coupling singularities when $A_1 \to \infty$ or $A_2
\to \infty$.  The $\bZ_2$-invariant couplings
\be
g_1 \equiv {A_2\over A_1^2} \qquad {\rm and} \qquad
g_2 \equiv {A_1^2\over A_2^2},
\label{gsum}
\ee
are related to the $\{\tau_1,\tau_2\}$ gauge couplings of the $SU(2)$
and $Sp(2)$ factors, respectively, by $\{g_1,g_2\}\sim \{
e^{i\pi\tau_1}, e^{i\pi\tau_2} \}$ at weak coupling.

\subsubsection{$\msusu \leftrightarrow \msusp$ map}

Finally, the low energy $SU(2)\times SU(2)$ and $SU(2)\times
Sp(2)$ descriptions of the theory are found to be equivalent
as long as the parameters of one theory are mapped to those
of the other by $\T: \msusp \to \msusu$ defined by
\be\label{Tmap}
\pmatrix{B_1\cr B_2\cr} = \T \pmatrix{A_1\cr A_2\cr}
\equiv \pmatrix{(A_1-8)/(4\alpha)\cr 
(A_2-6A_1+18)/(16\alpha^2)\cr}
\ee
where
\be\label{Tmap2}
16\alpha^3 = 2 A_1 - A_2 - 2,
\ee
or its inverse
\be\label{Tinv}
\pmatrix{A_1\cr A_2\cr} = \T^{-1} \pmatrix{B_1\cr B_2\cr}
\equiv \pmatrix{8 + 4\beta B_1\cr
30 + 24\beta B_1 + 16\beta^2 B_2\cr}
\ee
with
\be\label{Tinv2}
0 = 1 + \beta B_1 + \beta^2 B_2 + \beta^3 .
\ee


\section{S-duality in the SU(2)\,x\,SU(2) theory}

We will now derive the enlarged S-duality group of the $SU(2) \times
SU(2)$ theory.  The idea is a straightforward generalization of the
strategy used for a single $SU(2)$ factor reviewed in section 2: the
$SU(2)\times SU(2)$ model can be reached by flowing down from both the
$SU(n)\times SU(n)$ and $SU(2n)\times Sp(2n)$ series.  Denoting
by $\M$ the true coupling space of the $SU(2)\times SU(2)$ theory, we
therefore expect to find some multiple cover $\msusu$ of $\M$ as the
coupling space realized by flowing down in the $SU(n)\times SU(n)$
series, and a different multiple cover $\msusp$ by flowing down in the
$SU(2n)\times Sp(2n)$ series.  We then use the equivalence of the two
descriptions of the theory to deduce a map identifying $\msusu$ with
$\msusp$.  If this map is not a simple one-to-one map, then we thereby
deduce extra identifications leading to the ``smaller'' coupling space
$\M$ and therefore a larger S-duality group $\pi_1(\M)$.

The determination of $\msusu$ and $\msusp$ is easy, as we have
already done it in \cite{ab9804}.  There we showed that the embedding
argument leads to a coupling space for the $SU(n)\times SU(n)$ theory
which is precisely the $\msusu$ described above in
eqns.~(\ref{z3sum}--\ref{susings}), and likewise that the $SU(2n)
\times Sp(2n)$ theory coupling space is the $\msusp$ described above
in eqns.~(\ref{z2sum}--\ref{spsings}).  The map between $\msusu$ and
$\msusp$ is then the one derived at length in the last section, and
summarized in eqns.~(\ref{Tmap}--\ref{Tinv2}).  As this map is obviously
not one-to-one, we have therefore found new S-duality identifications
on the $SU(2) \times SU(2)$ coupling space, which is what we aimed to
show.

The remainder of this section will be devoted to understanding some
properties of the $\msusu \leftrightarrow \msusp$ map $\T$
(\ref{Tmap}--\ref{Tinv2}), and thereby of the the resulting enlarged
S-duality group, $\Gamma = \pi_1(\M)$.  We will refer to the $\bC^2$
of $B_i$ parameters as $\CB$ and of $A_i$ parameters $\CA$.  To see
algebraically the extra identifications induced on $\msusu$ by the map
$\T:\CA \to \CB$, we use it to construct maps from $\CB$ to
itself.  Note first that $\T$ and $\T^{-1}$ each have three image
points corresponding to the three different values that $\alpha$ or
$\beta$ can take.  In the case of $\T$, the three $\alpha$'s differ
only by cube root of unity phases, and the three image points in $\CB$
are related by the $\bZ_3$ identification $\P$ (\ref{z3sum}).  One the
other hand, the image of $\T^{-1}$ is generically three distinct
points in $\CA$ unrelated by the $\bZ_2$ identification $\R$
(\ref{z2sum}).  However, the images under $\T^{-1}$ of three points in
$\CB$ related by $\P$ are all the same three points in $\CA$, since a
$\P$ action on the $B_i$ just rotates the roots of (\ref{Tinv2}),
leaving (\ref{Tinv}) invariant.  

Since the $\T$ map commutes with $\P$, we can formulate the
identifications directly on $\Cf \equiv \CB/\{\P\}$ with coordinates
$f_i$ given by (\ref{fsum}).  In these variables $\T$ becomes
\be
\pmatrix{f_1 \cr f_2 \cr} = \T \pmatrix{A_1 \cr A_2\cr}
= \pmatrix{ 4(A_1-8)(2A_1-A_2-2)(A_2-6A_1+18)^{-2} \cr
(A_2-6A_1+18)(A_1-8)^{-2}\cr},
\ee
and $\T^{-1}$ reads
\be
\pmatrix{A_1 \cr A_2\cr} = \T^{-1} \pmatrix{f_1 \cr f_2 \cr}
= \pmatrix{ (8f_2+4\gamma)/f_2 \cr
(30f_2 + 24\gamma+16\gamma^2)/f_2 \cr},
\ee
where $\gamma$ is a root of
\be\label{gampam}
0 = f_1 \gamma^3 + \gamma^2 + \gamma + f_2 .
\ee
Note that while $\T$ is a single map, $\T^{-1}$ is generically three
maps, one for each $\gamma$ satisfying (\ref{gampam}).  Nevertheless,
it is easy to check that $\T \cdot \T^{-1}$ maps points in $\Cf$ to
themselves, and it follows that repeated applications of $\T$ and
$\T^{-1}$ generate no further identifications between $\CA$ and $\Cf$.

Since $\msusu \simeq \Cf/\{\Q\}$ and $\msusp \simeq \CA/\{\R\}$,
where $\Q$ and $\R$ are the $\bZ_2$ identifications $\Q:f_1 \leftrightarrow
f_2$, and $\R: A_1 \leftrightarrow -A_1$, further identifications will arise
upon combining $\T$ with $\R$ and $\Q$.  It is algebraically
easiest to work on $\Cf$ where there are three generators of non-trivial
maps involving $\T$, namely $\S_i \equiv \T \cdot \R \cdot \T^{-1}$.  
Explicitly, this map reads
\be\label{SmapS}
\S_i \pmatrix{f_1\cr f_2\cr} = \pmatrix{ (4f_2+\gamma)(3f_2+2\gamma
+\gamma^2) (6f_2+3\gamma+\gamma^2)^{-2} \cr
f_2 (6f_2+3\gamma + \gamma^2) (4f_2+\gamma)^{-2} \cr},
\ee
where $\gamma$ is a root of (\ref{gampam}).  The subscript on $\S$
denotes the three different choices of roots for $\gamma$, which lead
generically to three different image points in $\Cf$.  Thus the new
S-duality identifications $\S_i$ that we have found imply at least a
four-fold identification on $\Cf$ (the original point and its three
images).  Furthermore, the orbit of a given point under repeated
applications of the $\S_i$ can be shown to be just this set of four
points.  The $\bZ_2$ identification $\Q$ on $\Cf$ does not commute
with the $\S_i$, though it can be shown that for a given $\S_i$, there
exist an $\S_j$ and an $\S_k$ such that $\S_i \Q \S_j \Q \S_k \Q = 1$.
The minimum orbit of a generic point satisfying these relations
comprises 20 points, as shown in Fig.~1.  In fact this is the generic
orbit in $\Cf$ under the complete set of identifications generated by
$\S_i$ and $\Q$, as checked numerically.

\begin{figure}
\centerline{\psfig{figure=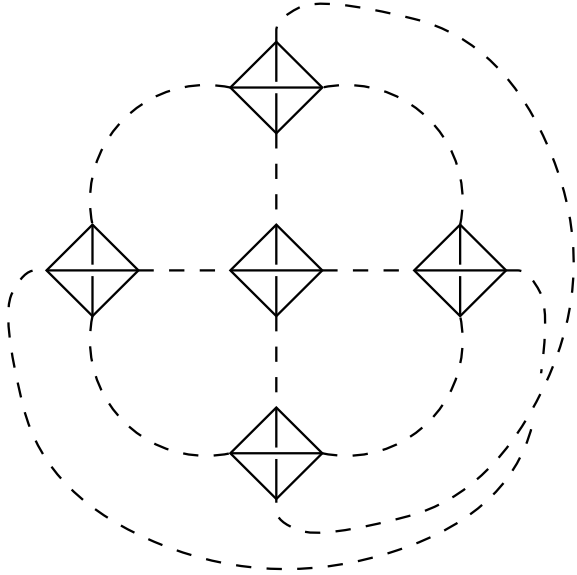,width=3truein}}
{\caption{The generic orbit of a point in $\Cf$ under $\{\S_j,\Q\}$.
The solid lines (edges of the tetrahedra) denote the action of the
$\S_j$ maps, while the dashed lines connecting the tetrahedra
denote the action of the $\Q$ map.}\label{fig1}}
\end{figure}

In summary, because of the algebraic complexity of the $\S_i$
generators, we have been unable to find a simpler description of the
resulting true coupling space $\M$ than
\be
\M \simeq \Cf/\{\Q,\S_i\}
\ee
with punctures at points satisfying (\ref{susings}) which reads in
the $f_i$ coordinates:
\be\label{fpunct}
0 = 1 - 4 f_1 - 4 f_2 + 18 f_1 f_2 - 27 f_1^2 f_2^2 ,
\ee
as well as weak coupling singularities when $f_1f_2=0$.
For clarity, we emphasize that $\pi_1(\M)$---the S-duality group of
$\M$---is {\it not} just the group generated by $\Q$ and $\S_i$.
There are many reasons for this: $\Cf$ already has $\bZ_3$ orbifold
points at $f_1 = f_2 = 0$ and $f_1 = f_2 = \infty$; $\Q$ and $\S_i$
act with fixed points; there are also strong and weak coupling
punctures on $\Cf$; finally, $\Q$ and $\S_i$ do not even generate a
group since there is no consistent labeling of the $\S_i$---the three
roots of (\ref{gampam})---on the whole of $\Cf$.

We can, however, argue that $\M$ is {\it not} just the Cartesian
product of two copies of the fundamental domain of $SL(2,\bZ)$ as one
might naively have guessed.  If it were this product, $\M$ would have
(complex) lines of $\bZ_3$ fixed points, whereas it is straightforward
to check that $\Q$ and $\S$ only have isolated $\bZ_3$ fixed points
which occur when $(f_1,f_2)$ is one of
\be\label{z3pntsM}
\left({1\over3},0\right), 
\left(0,{1\over3}\right), 
\left({1\over3},{1\over3}\right), 
\left({37+i\sqrt{3}\over98},{37-i\sqrt{3}\over98}\right),
\left({37-i\sqrt{3}\over98},{37+i\sqrt{3}\over98}\right).
\ee
In fact, these five points are all identified under $\Q$ and $\S_i$,
so there is only a single $\bZ_3$ fixed point in $\M$.

Note that the first two entries in (\ref{z3pntsM}) are the $\bZ_3$ points,
identified in section 2, on the coupling space of one $SU(2)$ factor in 
the limit where the other is decoupled.  The orbit of the $\bZ_2$ points
$f=2/9$, $\infty$, of a single $SU(2)$ factor also includes points at
strong coupling:
\be
\left({2\over9},0\right), 
\left(0,{2\over9}\right), 
\left(\infty,0\right), 
\left(0,\infty\right), 
\left({1\over3},{1\over3}\right), 
\left(-1,-1\right), 
\left({5\over16},{8\over25}\right),
\left({8\over25},{5\over16}\right).
\ee
In fact, there are whole (complex) lines of $\bZ_2$ fixed points.
Though they are hard to characterize explicitly, they all seem to be
images of the $\Q$ fixed line $f_1=f_2$ under $\S_i$.  These images
intersect in an $S_3$ orbifold point whose orbit in $\Cf$ is
\be
\left(\infty,\infty\right), 
\left({3\over8},{1\over3}\right), 
\left({1\over3},{3\over8}\right),
\left({1\over2},{1\over2}\right).
\ee

These examples illustrate the interesting feature of the $\Q$ and
$\S_i$ maps that they equate ``strong coupling'' punctures---$f_i$
satisfying (\ref{fpunct})---with weak coupling punctures satisfying
$f_1f_2=0$.  This is true in general: all strong coupling punctures
are so identified with weak coupling points.  To see this, note that
$f_1$ and $f_2$ satisfy (\ref{fpunct}) precisely when two roots of
(\ref{gampam}) coincide.  A double root of $\gamma$ satisfies
(\ref{gampam}) and its first derivative: $0 = 3 f_1 \gamma^2 + 2
\gamma + 1$.  Rewriting this as $f_1\gamma^3 = -(2 \gamma^2 +
\gamma )/3$ and substituting into (\ref{gampam}) gives $\gamma^2 + 2
\gamma + 3 f_2 = 0$.  But, by (\ref{SmapS}), this implies that
$\S_i(f_1)=0$ for this choice of the root $\gamma$, and thus that the
strong coupling puncture is mapped to a weak coupling singularity.
Thus S-duality identifications remove all ``ultra-strong'' coupling
points from $\M$, just as in the case of the $SL(2,\bZ)$ duality of a
single $SU(2)$ factor.

Also, the point $(1/3,1/3)$ is special as its image under $\S_i$
depends on how one approaches it.  Generically, its image is the point
$(1/3,0)$, but if one approaches it along the particular direction
$(f_1,f_2) = (1/3)\cdot (1 + \epsilon, 1 + \epsilon + k
\epsilon^{3/2})$, then its image under $S_i$ is the whole $(f,0)$
plane, where $f$ depends on $k$.

Finally, one should bear in mind that all our arguments only show that
the extra identifications on $\Cf$ leading to $\M$ are necessary, but
do not imply that there are no further identifications on $\M$.  In
principle one could rule out the existence of further identifications
from the low energy effective action on the Coulomb branch by showing
that the low energy data is different at distinct points of $\M$.
Though this is beyond the scope of the present paper, one piece of
evidence for there being no further identifications on $\M$ is the
fact, checked above (\ref{checkslz}), that in the limit where one of
the $SU(2)$ factors decouples, $\M$ already encodes the full
$SL(2,\bZ)$ S-duality group of the other $SU(2)$ factor.


\section*{Acknowledgments}
It is a pleasure to thank R. Maimon, J. Mannix, A. Shapere, G. Shiu,
H. Tye and P. Yi for helpful discussions.  This work is supported in
part by NSF grants PHY94-07194, PHY95-13717, and PHY97-22022.  The
work of PCA is supported in part by an A. P. Sloan Foundation
fellowship.

\end{document}